\documentclass[intlimits,twoside,a4paper]{article}

\usepackage{amsmath,amssymb}
\usepackage{graphicx}

\usepackage[T2A]{fontenc}
\usepackage[cp1251]{inputenc}

\usepackage{color}
\usepackage{gensymb}

\usepackage[eqsecnum]{cmpj2}

\issue{2016}{19}{1}{13004}
\doinumber{10.5488/CMP.19.13004}

\title[Cavity-ligand binding in a simple two-dimensional water model]%
{Cavity-ligand binding in a simple two-dimensional water model%
\thanks{Dedicated to the 65th birthday of Prof. Dr. Stefan Soko\l owski.}}
\author[G. Mazovec, M. Luk\v{s}i\v{c}, B. Hribar-Lee]{G. Mazovec, M. Luk\v{s}i\v{c},
B. Hribar-Lee}
\address{University of Ljubljana, Faculty of Chemistry and Chemical Technology, \\ Ve\v{c}na pot 113, SI-1000 Ljubljana, Slovenia}

\date{Received November 14, 2015, in final form December 22, 2015}
\authorcopyright{G. Mazovec, M. Luk\v{s}i\v{c}, B. Hribar-Lee, 2016}

\begin{document}

\maketitle

\begin{abstract}
By means of Monte Carlo computer simulations in the isothermal-isobaric ensemble, we investigated the interaction of a hydrophobic ligand with the hydrophobic surfaces of various curvatures (planar, convex and concave). A simple two-dimensional model of water, hydrophobic ligand and surface was used. Hydration/de\-hidration phenomena concerning water molecules confined close to the molecular surface were investigated. A notable dewetting of the hydrophobic surfaces was observed together with the reorientation of the water molecules close to the surface. The hydrogen bonding network was formed to accommodate cavities next to the surfaces as well as beyond the first hydration shell. The effects were most strongly pronounced in the case of concave surfaces having large curvature. This simplified model can be further used to evaluate the thermodynamic fingerprint of the docking of hydrophobic ligands.
\pacs 02.70.Uu, 05.10.Ln, 61.20.Gy, 61.20.Ja, 61.30.Hn
\end{abstract}

\section{Introduction}

Due to the important influence of the confinement on the structural,
thermodynamical, and phase equilibrium properties of
simple fluids, these systems have been extensively studied recently \cite{Rice2014,Pizio2004,Pizio2007}. The topic is also of great interest
in the studies of biological systems in which many unanswered questions are concerned with hydration
and dehydration of molecules of arbitrary shapes where water as a solvent finds itself confined close to the
molecular surface.

One of the important issues that have begun to emerge is the
effect of the shape, curvature, and roughness of a surface on its
interaction with a ligand \cite{Matmakulov2004,Rudich2000,Southall2002,Wallqvist1995,Pratt1980,Chorny2005,Baron2010}. In order to understand the phenomena such as binding of the ligand to the receptor, or
drugs to proteins, it is crucial to know the potential of the mean force
between two biomolecules in question. Virtually all the binding sites
in biology have a concave shape which imposes very particular geometrical
constraints to the solvated water \cite{Sharp1991} and, therefore,
the findings referring to the potential of the mean force between two
spherical surfaces cannot be generalized to these problems.

To summarize the recent molecular dynamics studies on model systems of
purely hydrophobic cavities
\cite{Baron2010,Baron2012,Baron2012a,Setny2010}, it has been
shown that water appears to be an active component in cavity-ligand
association. An important impact of changes in water structure during
the binding process has been noticed \cite{Baron2010,Setny2010}. While
the cavity hydration can be altered by changing its radius, in
weakly hydrated cavity regions the reorganization of water molecules
and suppression of the solvent
fluctuations leads to enthalpy driven association \cite{Baron2011}. A
different interpretation of the computer simulation results have been
given by Graziano \cite{Graziano2012}. His analysis shows that the Gibbs free energy gain upon
association of a ligand in a concave hydrophobic cavity is mainly due to the decrease in the solvent-excluded
volume, that translates in a gain of configurational-translational
entropy of water molecules. This entropic driving force is masked
by a large enthalpy gain associated with the reorganization of
water-water hydrogen bonds upon association of the two nonpolar objects \cite{Graziano2012}.

In this work we have focused on the investigation of the potential of
the mean force between a hydrophobic ligand and a planar, concave, and
convex purely hydrophobic surface with different curvatures. Our main interest was in the
interpretation of the potential of the mean force through the water
microstructural changes due to the confinement. To better visualize these effects we have
chosen a simple two-dimensional Mercedes-Benz (MB) water model \cite{Dill2005}.

The paper is organized as follows: After this short introduction, the
model and method are described. Next, the results are presented and
discussed, and the conclusions are given in the end.

\section{Model and method}

The Mercedes-Benz model \cite{Dill2005} was used
to describe water molecules. In this model, the
water mole\-cule is represented as a two-dimensional Lennard-Jones (LJ)
disk with three equally separated hydrogen bonding arms, and interacts with another water molecule through the potential $U_\text{ww}$:
\begin{equation}
\label{eq:potvoda}
U_\text{ww}(\mathbf{X}_i,\mathbf{X}_j)=U_\textrm{LJ}(r_{ij})
+U_\textrm{HB}(\mathbf{X}_i,\mathbf{X}_j).
\end{equation}
$\mathbf{X}_i$ denotes the vector with coordinates and orientation of the
$i$-th particle, and
$r_{ij}$ is the distance between the centres of molecules $i$ and $j$.
Lennard-Jones 12-6 potential is:
\begin{equation}
\label{eq:LJ}
U_\textrm{LJ}(r_{ij})=4\epsilon_\textrm{LJ}\left[\left(\frac{\sigma_\textrm{LJ}}{r_{ij}}\right)^{12
}-\left(\frac{\sigma_\textrm{LJ}}{r_{ij}}\right)^6\right],
\end{equation}
where $\epsilon_\textrm{LJ}$ represents the depth of the potential well, and
$\sigma_\textrm{LJ}$ is the Lennard-Jones diameter.

The hydrogen bond strength is a Gaussian
function of the intermolecular distance and the angle between two hydrogen
bonding arms:
\begin{equation}
\label{eq:HB}
U_\textrm{HB}(\mathbf{X}_i,\mathbf{X}_j)=\epsilon_\textrm{HB}G(r_{ij}-r_\textrm{HB})\sum_{k,l=1}^3
G(\hat{\mathbf{{i}}}_k\cdot\hat{\mathbf{{u}}}_{ij}-1)G(\hat{\mathbf{{j}}}
_l\cdot\hat{\mathbf{{u}}}_{ij}+1),
\end{equation}
where $G(x)$ is an unnormalized Gaussian function:
\begin{equation}
\label{eq:gauss}
G(x)=\exp\left(-x^2/2\sigma^2\right).
\end{equation}
The unit vector $\hat{\mathbf{{i}}}_k$ represents the $k$-th arm of the $i$-th molecule
($k=1, \,2,\, 3$). Unit vector $\hat{\mathbf{u}}_{ij}$ is the
direction vector of the line joining the centres of the $i$-th and $j$-th molecule.
Parameters $\epsilon_\textrm{HB}=-1$ and $r_\textrm{HB}=1$ determine the energy and the
length of the optimal hydrogen bond. Values for the model parameters are as follows: $\sigma_\textrm{LJ}=0.7 \, r_\textrm{HB}=0.7$,
$\epsilon_\textrm{LJ}=0.1\left|\epsilon_\textrm{HB}\right|=0.1$, and $\sigma=0.085$.

The hydrophobic ligand was represented as a Lennard-Jones disk of the same size ($\sigma_\textrm{LJ}=0.7$) as the water molecule.

\begin{figure}[!t]
\centering
\includegraphics[width=6.5cm]{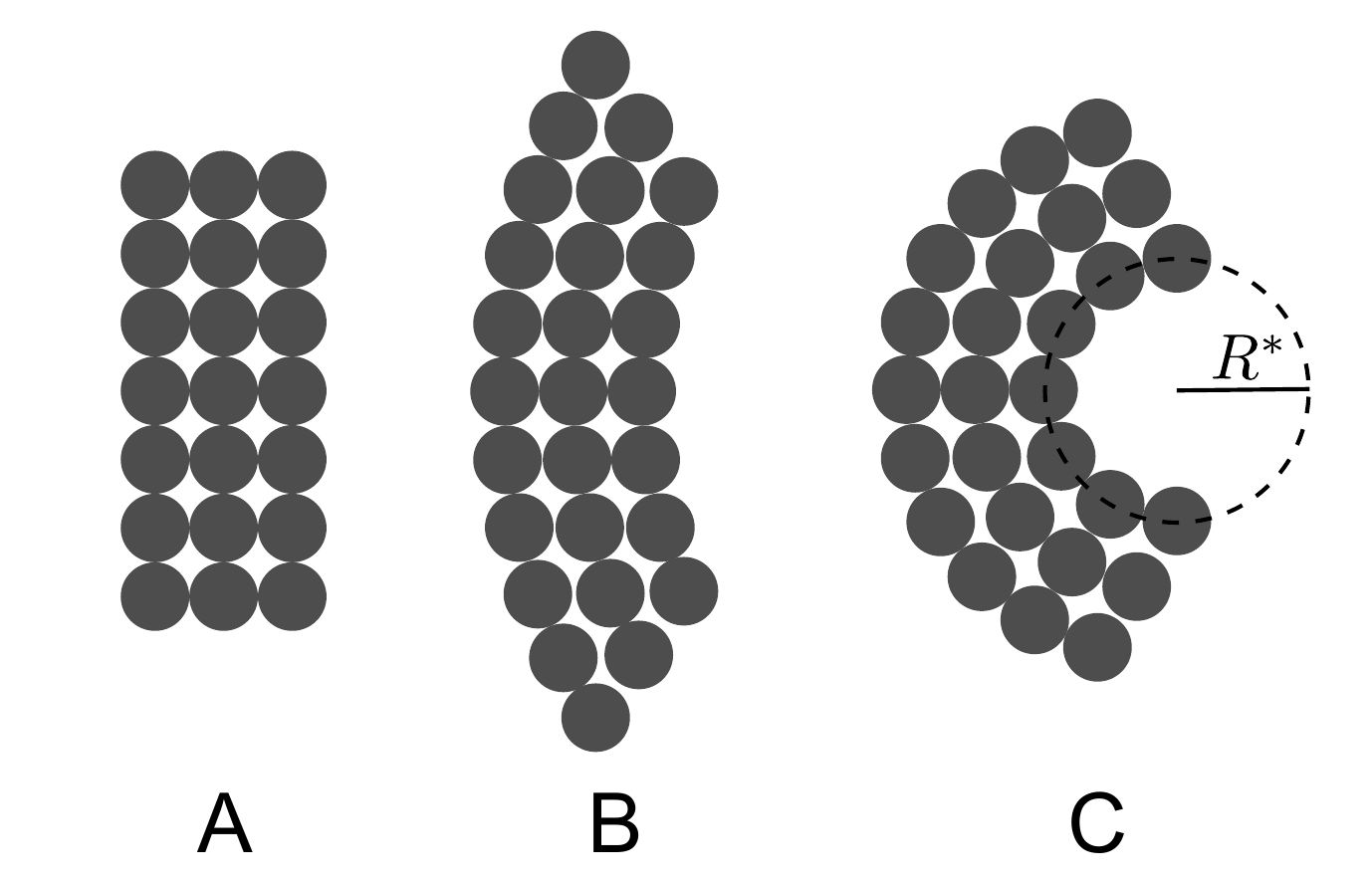}
\caption{Schematic representation of the surfaces used in simulations. All surfaces were made of three layers of hydrophobic particles with Lennard-Jones (LJ) diameter $\sigma_\textrm{LJ} = 0.7$. Planar surface (A) was composed of $3\cdot7=21$ LJ discs, while curved surfaces (B and C) were made of $7+9+11=27$ LJ discs. Radii of the inner and outer layer of the surface B were $R^*=4.55$ and 5.95, respectively, while for the surface C the inner and the outer radii were 1.40 and 2.80, respectively. The surfaces were placed in the middle of the simulation box.}
\label{fig:surfaces}
\end{figure}

To model a hydrophobic surface of a given curvature, a molecular wall
was formed from Lennard-Jones discs (see figure~\ref{fig:surfaces}). One surface was planar, while the others were bent. The interaction of the test water or hydrophobic ligand with the surfaces having positive curvatures (concave), and with negative curvatures (convex) was tested. The two radii used for concave surfaces were $R^* = 4.55$ and 1.40, while for the convex surfaces the radii were 2.80 and 5.95. The hydrophobic particle and the water molecules interacted with the particles forming the hydrophobic surface through the Lennard-Jones potential [equation (\ref{eq:LJ})]. The LJ parameters ($\sigma_\textrm{LJ}$ and $\epsilon_\textrm{LJ}$) were the same for all particles in question.

Reduced units were used throughout the paper: $r^* = r/r_\textrm{HB}$, $T^* = k_\textrm{B}T/|\epsilon_\textrm{HB}|$, $p^* = pr_\textrm{HB}^2/|\epsilon_\textrm{HB}|$.

Isobaric-isothermal Monte Carlo computer simulations \cite{CSL} were performed for systems with 120 MB water molecules and a single hydrophobic ligand (test particle). The hydrophobic surface was placed in the centre of the simulation box. Reduced temperature, $
T^*$, and pressure, $p^*$, of the systems were 0.20 and 0.19, respectively.
Initial configuration of water molecules was
chosen randomly and then equilibrated in $10^7$ cycles long simulation.
Statistics were then collected over $10^8$ cycles long production run which
started from equilibrated configuration. In one cycle, either rotation or displacement (probabilities for rotation and displacement being
equal) of a water molecule was attempted. Maximal displacement and rotation were adjusted throughout the simulation, so that approximately one half of Monte Carlo moves were accepted. After every 120th cycle, an attempt to change the volume of the simulation box was made, where the maximal change was adjusted in
the same way as maximal displacement/rotation.

The pair correlation and angular distribution functions of water next
to the surface were calculated using the histogram method, while the
potential of the mean force between the hydrophobic ligand and the surface was
calculated using the Widom's insertion technique \cite{Widom}.

\section{Results and discussion}

\begin{figure}[!b]
\centering
\includegraphics[width=14cm]{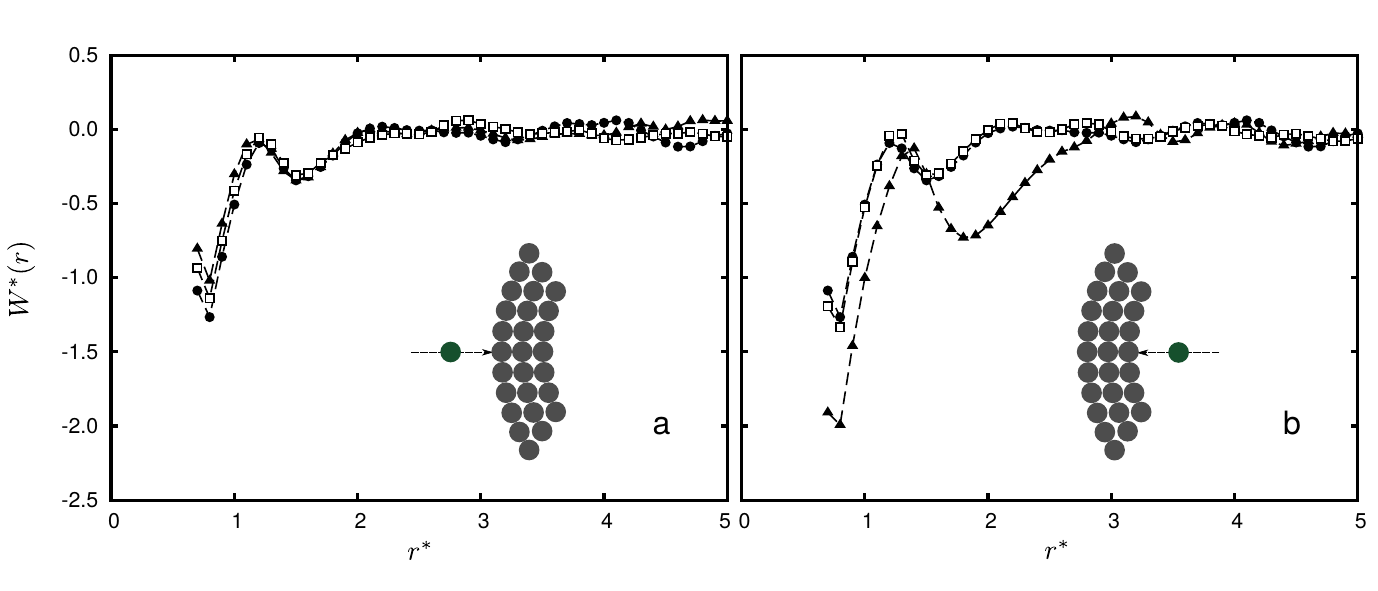}
\vspace{-3mm}
\caption{Potentials of the mean force (PMFs) between a hydrophobic ligand (test particle) and the central particle of the hydrophobic surface. Centre-to-centre distance between test particles is given by $r^*$ (in the direction of the $x$-axis; see the insets). Results for the convex shaped surfaces are shown in panel~(a), and for the concave case in panel~(b). The reduced PMF, $W^*(r) = W(r)/k_\text{B}T$, for the case of linear surface is shown in both panels (circles). Two curvatures of the surface were tested~--- panel~(a): 5.95 (squares) and 2.80 (triangles), panel~(b): 4.55 (squares) and 1.40 (triangles). $T^* = 0.20$ and $p^*=0.19$.}
\label{fig:Wr}
\end{figure}

Here, we present the results of the Monte Carlo simulations of
interaction of the hydrophobic ligand with the hydrophobic surface in
water. All the results are given for $T^*=0.2$ and $p^*=0.19$. Figure~\ref{fig:Wr}~(a) shows the potentials of the mean force (PMFs) between the
hydrophobic particle and the central particle of the surface for
convex surfaces, while the PMFs for concave surfaces are shown in figure~\ref{fig:Wr}~(b). The PMF with planar surface is shown for comparison in both panels. An important difference from the PMF between two hydrophobic discs in water (figure~1 of reference \cite{Southall2002}) is observed. There is a larger difference in the values of the peaks belonging to the contact minimum (CM) and solvent separated minimum (SSM) compared to the homogeneous system (two hydrophobic discs); CM state is here much more stable than the SSM state. This suggests that the hydrophobic particle would prefer to stay close to the surface. Not much difference between the PMFs for different surface curvatures is observed in the convex cases, while in the case of strongly concave surface, the contact state is additionally stabilized and the position of the SSM is shifted further away from the surface.

 \begin{figure}[!t]
\centering
\includegraphics[width=14cm]{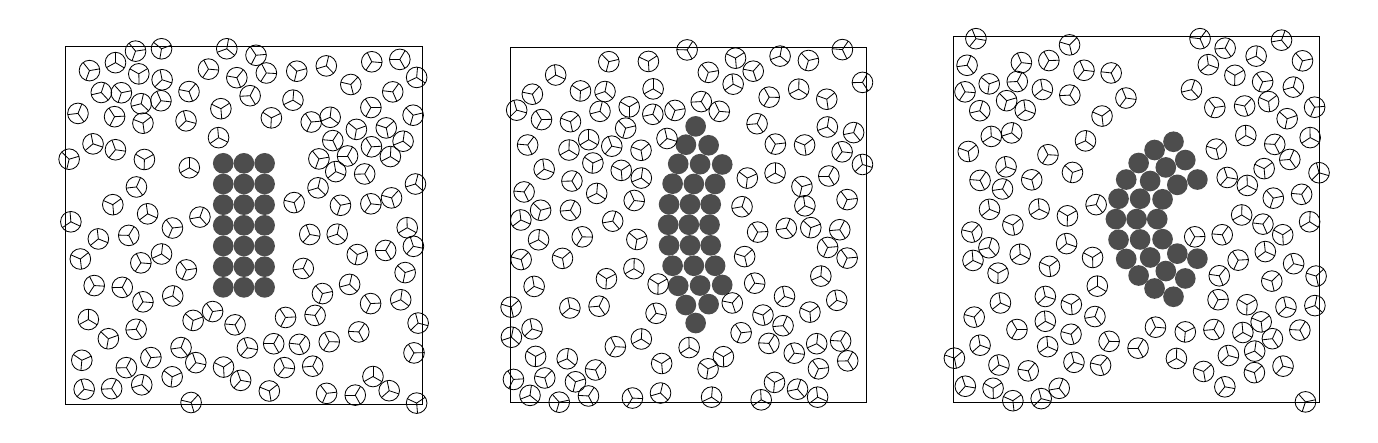}
\caption{Representative simulation snapshots for various surfaces. White circles represent MB water molecules with three hydrogen bonding arms (Mercedes-Benz logo). The grey circles are hydrophobic particles forming the surfaces. $T^* = 0.20$ and $p^*=0.19$.}
\label{fig:snap}
\end{figure}

\begin{figure}[!b]
\centering
\vspace{-4mm}
\includegraphics[width=14cm]{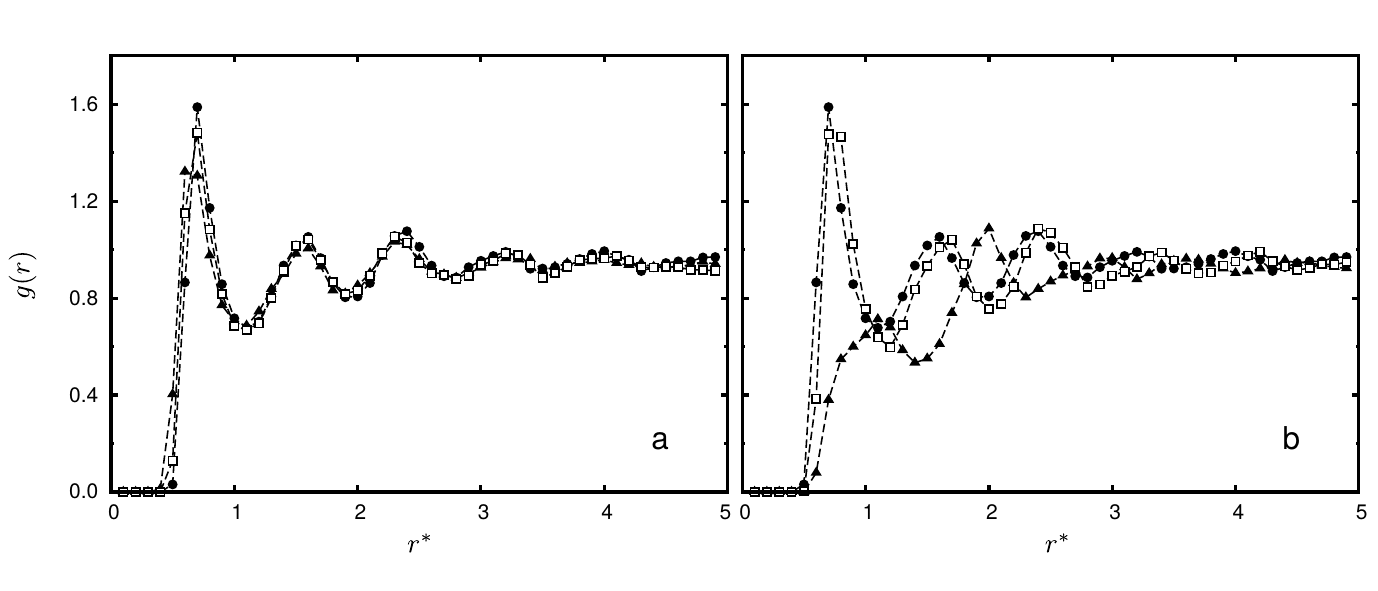}
\vspace{-4mm}
\caption{Radial distribution functions between a test water molecule and the central particle of the hydrophobic surface. Distance $r^*$ has the same meaning as in figure~\ref{fig:Wr}. Surface parameters, notations and conditions ($T^*,p^*$) are the same as in figure~\ref{fig:Wr}.}
\label{fig:gr}
\end{figure}

To interpret these results in view of water microstructure, we analysed the hydration of the surfaces more in detail. Characteristic simulation snapshots are shown in figure~\ref{fig:snap}. A dehydration of the hydrophobic surface is noticed for all surfaces studied. The water molecules are pushed away and cavities are formed at the surface contact as well as beyond the first layer of water molecules. This is further confirmed by the water-surface radial distribution functions shown in figure~\ref{fig:gr}. All $g(r)$ show a layered structure of water molecules away from the surface. There is no significant qualitative difference in the $g(r)$ between the planar surface case and convex bent surfaces. The value of the contact peak slightly decreases with an increasing curvature [see panel~(a) of figure~\ref{fig:gr}].

On the other hand, a qualitatively different behaviour is observed in the case of strongly concave surface. Here, water is completely pushed away from the cavity, forming the first hydration layer approximately two hydrogen bonds away from the surface [figure~\ref{fig:gr}~(b)]. This is in agreement with the previous studies by Baron et al. and Graziano \cite{Baron2010,Baron2012,Baron2012a,Baron2011,Graziano2012}. Due to the incapability of forming hydrogen bonds, a water molecule in such a state is entropically stabilized.

\begin{figure}[!t]
\centering
\includegraphics[width=14cm]{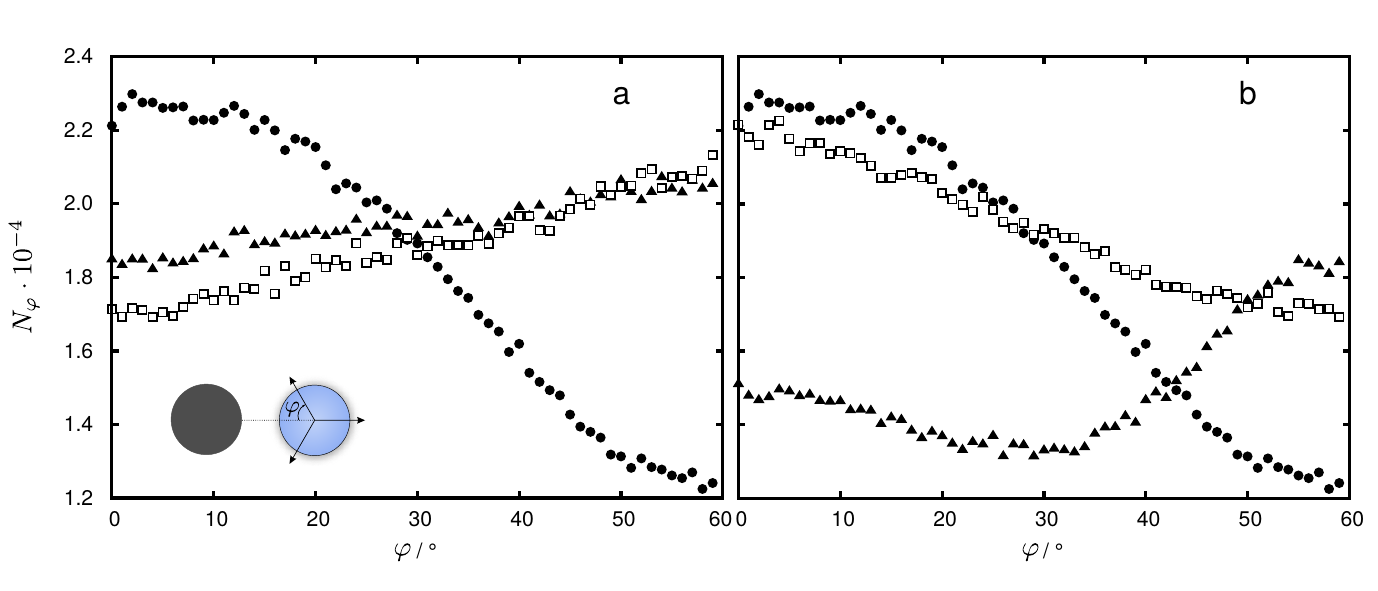}
\vspace{-3mm}
\caption{Angular distribution of the first hydration shell water with respect to the hydrophobic surface. Central particle of the surface closest to the test water was chosen to gather statistics. The angle $\phi$ is defined $0 \, \degree$ when one of the water's hydrogen bonding arms points towards the centre of the surface's particle [see the sketch in panel~(a)]. Results for the convex shaped surfaces are shown in panel~(a), and for the concave case in panel~(b). For comparison, the unnormalized distribution, $N_\varphi$, for the case of linear surface is shown in both panels (circles). Surface parameters, notations and conditions ($T^*,p^*$) are the same as in figure~\ref{fig:Wr}.}
\label{fig:angular}
\end{figure}

Further, the orientation of the surface water molecules belonging to
the first hydration shell was examined. Figure~\ref{fig:angular}~(a)
shows the results for the convex surfaces, and panel~(b) for the concave
surfaces. The case of planar wall is shown in both panels for
comparison. As already noticed by Southall et al. \cite{Southall2000},
a model water molecule close to the planar surface can no longer form
all three hydrogen bonds with neighbouring water molecules. To
compensate for this loss, it ``wastes'' one of the hydrogen bonds by
pointing it towards the surface. This is seen from the angular
distribution where the most probable orientation of the water molecule
at the planar surface is the one pointing the hydrogen bond towards
the surface ($\varphi = 0\, \degree$). The two hydrogen bonds pointing away from
the surface participate in the formation of cavities facilitating
 a favourable SSM in the PMFs.

As the surface adopts the convex shape, the
angular preference of the water to the surface is much less
expressed. For the case in panel~(a) of figure~\ref{fig:angular}, we can conclude that angular
preferences of  water have a negligible contribution to the shape of the PMFs.

The situation gets reversed in the case of strongly concave surface
[triangles in figure~\ref{fig:angular}~(b)]. The preferential angle of water
molecule in this case is the one where the hydrogen bonding arm
points at an angle $60 \, \degree$ with respect to
the centre-centre coordinate [see the insert in panel~(a)], suggesting
the strongly obstructed hydrogen bond formation due to the lack of space in
the cavity. This way a water molecule was stabilized by an increase of the rotational entropy.

\section{Conclusions}

The results for the potential of the mean force between hydrophobic solute
and hydrophobic surface of various shapes in the model water solutions were
presented. The results were analysed in view of hydration water structure and orientation. All the hydrophobic surfaces studied showed dewetting which was most pronounced for the bent concave surface. The angular orientation of the water molecules facilitated the formation of cavities which stabilized the non-covalent direct and water separated binding of the hydrophobic molecule to the surface. In the future work, temperature dependence of the phenomena will be investigated and the results will be used for the enthalpy-entropy decomposition of the free energy of ligand-surface interaction.

\clearpage

\section*{Acknowledgements}
This work was supported by the Slovenian Research Agency through
grant P1-0103-0201 and Slovene-Korean bilateral grant BI-KR/13-14-003.
M.L. and B.H.-L. acknowledge the
support of the NIH Grant \linebreak 2R01GM063592-14.

\ukrainianpart

\title{Зв'язки порожнина-ліганд у простій двовимірній \\ моделі води}
\author{Г. Мазовець, М. Лукшич, Б. Грібар-Лі}
\address{Університет Любляни, факультет хімії і хімічної технології, Любляна, Словенія}

\makeukrtitle

\begin{abstract}
\tolerance=3000%
Використовуючи комп'ютерне моделювання в ізотермічно-ізобаричному ансамблі, досліджено взаємодію гідрофобного ліганда  з гідрофобними поверхнями різної кривизни (плоскі, випуклі та вгнуті). Використано просту двовимірну модель води, гідрофобний ліганд та гідрофобну поверхню.
Досліджено явища гідратації/дегідратації молекул води близьких до молекулярної поверхні. Спостерігалось суттєве незмочування гідрофобних поверхонь поряд  з переорієнтацією молекул води поблизу поверхні. Спостережено формування мережі водневих зв'язків так що
порожнини розташовуються біля поверхні, а також поза межами  першої гідратної оболонки.
 Згадані ефекти найбільш виражені у випадку ввігнутих поверхонь з великою кривизною. Цю спрощену модель можна використовувати щоб оцінити термодинамічні аспекти докування гідрофобних лігандів.
\keywords зв'язки порожнина-ліганд, утримання води, поверхнева гідратація, потенціал середньої сили, комп'ютерне моделювання Монте Карло, двовимірна модель води
\end{abstract}

\end{document}